%% file: ms.tex
\documentclass[apj]{emulateapj}

\usepackage{apjfonts,ulem,color,hyperref}
\usepackage{amssymb, amsmath}
\input{kbsmacs.tex}
\shorttitle{A Search for Water in the Atmosphere of HAT-P-26\lowercase{b} Using LDSS-3C}
\shortauthors{Stevenson {\em et al.}}

\submitted{}

\begin{document}

\title{A Search for Water in the Atmosphere of HAT-P-26\lowercase{b} Using LDSS-3C}

\author{Kevin B.\ Stevenson\altaffilmark{1,5}}
\author{Jacob L.\ Bean\altaffilmark{1}}
\author{Andreas Seifahrt\altaffilmark{1}}
\author{Gregory J.\ Gilbert\altaffilmark{1}}
\author{Michael R.\ Line\altaffilmark{2}}
\author{Jean-Michel~D\'esert\altaffilmark{3}}
\author{Jonathan J.\ Fortney\altaffilmark{4}}
\affil{\sp{1} Department of Astronomy and Astrophysics, University of Chicago, 5640 S Ellis Ave, Chicago, IL 60637, USA}
\affil{\sp{2} NASA Ames Research Center, Moffett Field, CA 94035}
\affil{\sp{3} Anton Pannekoek Institute for Astronomy, University of Amsterdam, The Netherlands}
\affil{\sp{4} Department of Astronomy and Astrophysics, University of California, Santa Cruz, CA 95064}
\affil{\sp{5} NASA Sagan Fellow}

\email{E-mail: kbs@uchicago.edu}

\begin{abstract}
The characterization of a physically-diverse set of transiting exoplanets is an important and necessary step towards establishing the physical properties linked to the production of obscuring clouds or hazes.  It is those planets with identifiable spectroscopic features that can most effectively enhance our understanding of atmospheric chemistry and metallicity.  The newly-commissioned LDSS-3C instrument on Magellan provides enhanced sensitivity and suppressed fringing in the red optical, thus advancing the search for the spectroscopic signature of water in exoplanetary atmospheres from the ground.  Using data acquired by LDSS-3C and the {\em Spitzer Space Telescope}, we search for evidence of water vapor in the transmission spectrum of the Neptune-mass planet HAT-P-26b.  Our measured spectrum is best explained by the presence of water vapor, a lack of potassium, and either a high-metallicity, cloud-free atmosphere or a solar-metallicity atmosphere with a cloud deck at $\sim$10~mbar.  The emergence of multi-scale-height spectral features in our data suggests that future observations at higher precision could break this degeneracy and reveal the planet's atmospheric chemical abundances.  We also update HAT-P-26b's transit ephemeris, $t$\sb{0}~= 2455304.65218(25)~BJD\sb{TDB}, and orbital period, $p$~= 4.2345023(7)~days.
\end{abstract}

%%%%%%%%%%%%%%%%%%%%%%%%%%%%%%%%%%%%%%%%%%%%%%%%%%%%%%%%%%%%%%%%%%%%%%%%%%%%%%%
\section{Introduction}
\label{intro}
%%%%%%%%%%%%%%%%%%%%%%%%%%%%%%%%%%%%%%%%%%%%%%%%%%%%%%%%%%%%%%%%%%%%%%%%%%%%%%%

Understanding the prevalence of clouds and hazes is one of the major outstanding issues in exoplanetary atmospheres \citep[e.g.,][]{Pont2013}.  As more planets are being studied with space-based and large ground-based telescopes, the publication of featureless transmission spectra is becoming a more common occurrence \citep[e.g.,][]{Gibson2013b, Kreidberg2014, Knutson2014, Mallonn2015a}.  Correspondingly, the number of detections of spectroscopic features is also on the rise \citep[e.g.,][]{Deming2013, Huitson2013, Mandell2013, Fraine2014, McCullough2014, Kreidberg2014b, Kreidberg2015, Wilson2015}.  Although temperature and surface gravity are thought to play a significant role in the production of clouds and hazes in exoplanet atmospheres \citep[e.g., ][]{Kempton2012, Helling2013, Morley2013, Morley2015}, work is still ongoing to establish trends in the available data.  To advance our knowledge, we need to obtain precise transmission spectra of targets that have a variety of physical characteristics and continue investigating potential correlations.

HAT-P-26b is an inflated, Neptune-mass planet that orbits its K1 host star every 4.23 days \citep{Hartman2011}.  As a result, the planet has a relatively low surface gravity ($\log(g) = 2.66$~dex, $g = 4.55$~m~s\sp{-2}) that is well-suited for atmospheric characterization and an equilibrium temperature ($T$\sb{eq}~=~990~K) that may allow for the detection of both water and methane.  To date, there are no previously-published results detailing the atmospheric characterization of HAT-P-26b.

The layout of this manuscript is as follows.  Section \ref{sec:ldss} introduces the new Low Dispersion Survey Spectrograph 3C (LDSS-3C) instrument.  In Section \ref{sec:obs}, we discuss the acquisition and reduction of LDSS-3C and {\em Spitzer} data, the handling of position- and time-dependent systematics in our light-curve model fits, and the estimation of parameter uncertainties.  Section \ref{sec:results} presents new constraints on the planet's orbital period, matches our measured transmission spectrum to representative atmospheric models, and compares HAT-P-26b to other well-characterized exoplanets.  Finally, we offer our conclusions in Section \ref{sec:concl}.

%%%%%%%%%%%%%%%%%%%%%%%%%%%%%%%%%%%%%%%%%%%%%%%%%%%%%%%%%%%%%%%%%%%%%%%%%%%%%%%
\section{An Introduction To LDSS-3C}
\label{sec:ldss}
%%%%%%%%%%%%%%%%%%%%%%%%%%%%%%%%%%%%%%%%%%%%%%%%%%%%%%%%%%%%%%%%%%%%%%%%%%%%%%%

The Low Dispersion Survey Spectrograph (LDSS) is an optical imaging spectrograph with multi-object capabilities mounted on the 6.5 meter Magellan II (Clay) Telescope at Las Campanas Observatory (LCO) in Chile.  \footnote{More information on observing with LDSS-3C can be found on the instrument webpage for Las Campanas Observatory: \href{http://www.lco.cl/telescopes-information/magellan/instruments/ldss-3/ldss3_c}{\color{blue}www.lco.cl/telescopes-information/magellan/instruments/ldss-3/ldss3\_c}}  It operates as a wide-field imager or as a multi-object spectrograph with up to eight custom multi-aperture masks held in a wheel. The instrument uses all refractive optics and focuses the light on an external detector at a focal ratio of f/2.5.  First used at Magellan in 2001, LDSS underwent a major upgrade of its optics, including new VPH grisms, in 2005. At this time a thinned, back-illuminated STA0500A 4k$\times$4k CCD was installed and the instrument was dubbed LDSS-3. 

In order to fully utilize the excellent red optical and near infrared throughput of the LDSS-3 optics, the detector system of LDSS-3 was upgraded in a joint effort between the Fermi National Accelerator Laboratory (Fermilab) and the University of Chicago.  The new detector is a CCD left over from the Dark Energy Camera production \citep{Flaugher15}.  The CCD is a fully depleted, 250\,{\micron} thick detector with superior quantum efficiency in the near infrared compared to the much thinner STA0500A CCD. These types of detectors were developed by Lawrence Berkeley National Laboratory (LBNL) using high resistivity Si substrates. For the DECam project, LBNL supplied bare CCD dies and Fermilab packaged them into custom back-side-illuminated modules \citep{Derylo06,Estrada10}. A custom CCD dewar was designed and built at the Exoplanet Instrument Lab of the University of Chicago as a seamless replacement for the old detector system.  The CCD was fully characterized at Fermilab. The integrated detector system, including the readout electronics, was then thoroughly tested at the University of Chicago. The new detector system was installed and commissioned in March 2014 on LDSS-3 at LCO.  The newly upgraded instrument is now dubbed LDSS-3C.  In this section we report on the results from the detector characterization in the lab and from the commissioning observations obtained 2014 March 21 -- 24.  Table \ref{tab:ldss3c} provides additional information on the detector.

\begin{deluxetable}{llll}
\tablecolumns{4}
\tablewidth{0pc}
\tablecaption{\label{tab:ldss3c}
Properties of LDSS-3C}
\tablehead{
\colhead{Property}    &  \multicolumn{3}{c}{Value}  \\ 
}
\startdata
\multicolumn{1}{l}{Active area} & \multicolumn{3}{l}{2048 $\times$ 4096 pixel} \\
\multicolumn{1}{l}{Pixel size} & \multicolumn{3}{l}{15 $\mu$m} \\
\multicolumn{1}{l}{Quantum efficiency} & \multicolumn{3}{l}{68\% at 500\,nm} \\
\multicolumn{1}{l}{                               } & \multicolumn{3}{l}{94\% at 700\,nm}\\
\multicolumn{1}{l}{                               } & \multicolumn{3}{l}{48\% at 1000\,nm}\\
\multicolumn{1}{l}{Full well depth} & \multicolumn{3}{l}{$\ge$200,000\,e$^{-}$} \\
\multicolumn{1}{l}{1\% linearity limit} & \multicolumn{3}{l}{$\ge$175,000\,e$^{-}$} \\
\multicolumn{1}{l}{Damage threshold\footnote{Irreversible loss of full well depth and altered chip characteristics might occur when exceeding this level of saturation.} }& \multicolumn{3}{l}{2,000,000\,e$^{-}$\,s$^{-1}$} \\    
\multicolumn{1}{l}{ADC depth} & \multicolumn{3}{l}{16 bit} \\                        
\multicolumn{1}{l}{Dark current\footnote{At detector operating temperature of 173\,K.}}   & \multicolumn{3}{l}{35\,e$^{-}$\,hr$^{-1}$} \\
\multicolumn{1}{l}{CTI\footnote{Charge Transfer Inefficiency for left and right amplifier, respectively.}} & \multicolumn{3}{l}{5.5$\times 10^{-6}$ / 2.5$\times 10^{-5}$ (horizontal)} \\
\multicolumn{1}{l}{                               }  & \multicolumn{3}{l}{5.6$\times 10^{-6}$ / 1.1$\times 10^{-5}$ (vertical)} \\
\multicolumn{1}{l}{Amplifier crosstalk} & \multicolumn{3}{l}{$\le$0.02\% (for unsaturated sources)} \\
\multicolumn{1}{l}{                              }& \multicolumn{3}{l}{20--30\,e$^{-}$ (for saturated sources)}\\

\cutinhead{Readout characteristics}            
Read mode & Turbo & Fast & Slow \\
Readout speed (s) & 24 & 28 & 152 \\
Read noise\footnote{Practically achievable noise levels without electronic interference in parenthesis.} (e$^{-}$) & 10 (8) & 12 (5.5) & 5 (3) \\
Inv. gain\footnote{For left and right amplifier, respectively.} (e$^-$\,ADU$^{-1}$) & 2.7 / 3.1 & 1.5 / 1.8 & 0.16 / 0.19 \\

\cutinhead{Normal orientation}
\multicolumn{1}{l}{Spatial coverage\footnote{At field center.} } & \multicolumn{3}{l}{6.4\arcmin} \\
\multicolumn{1}{l}{Spectral coverage\footnotemark[6] } & VPH blue grism: &  400 -- 640 nm &\\
\multicolumn{1}{l}{}& VPH red grism: &  640 -- 1040 nm &\\
\multicolumn{1}{l}{}& VPH all grism: &  370 -- 1070 nm &\\

\cutinhead{``Nod \& Shuffle'' orientation}
\multicolumn{1}{l}{Spatial coverage\footnotemark[6] }& \multicolumn{3}{l}{8.3\arcmin} \\
\multicolumn{1}{l}{Spectral coverage\footnotemark[6] }& VPH blue grism: &  450 -- 588 nm &\\
\multicolumn{1}{l}{}& VPH red grism: &  720 -- 960 nm &\\
\multicolumn{1}{l}{}& VPH all grism: &  465 -- 880 nm &\\

\cutinhead{On sky characteristics}
\multicolumn{1}{l}{Imaging Field of View}& \multicolumn{3}{l}{8.3\arcmin\, diameter, mapped onto a 6.4\arcmin $\times$12.8\arcmin\, detector} \\
\multicolumn{1}{l}{Plate scale}& \multicolumn{3}{l}{0.189\arcsec/pixel} \\
Grisms & VPH red & VPH blue & VPH all\\
Central wavelength  &  850\,nm & 520\,nm & 650\,nm\\
Dispersion & 1.175\,\AA/pixel & 0.682\,\AA/pixel & 1.890\,\AA/pixel \\
Resolving power\footnote{For a 4 pixel (0.76\arcsec) wide slit.} & 1810 & 1900 & 860\\ 
Peak efficiency & 0.92 & 0.85 & ---\footnote{Currently not measured.}\\

\enddata
\end{deluxetable}

\begin{figure}[t]
\centering
\includegraphics[width=1.0\linewidth]{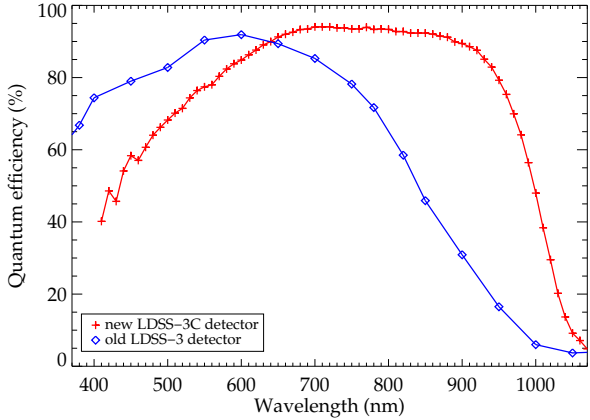}
\caption{\label{fig:QE}{
Quantum efficiency of the old 4k$\times$4k LDSS-3 CCD (blue diamonds) vs. the new fully depleted, 250\,{\micron} thick 2k$\times$4k CCD used for the upgrade to LDSS-3C (red crosses).
}}
\end{figure}

The new CCD, having the same pixel size (15\,{\micron}) as the old device but offering only 2048$\times$4096 unbinned pixels,  underfills the 8.3{\arcmin} diameter FOV of LDSS-3 by 10.5\%. In its standard orientation, the dispersion axis of the spectrograph is aligned with the columns of the detector, slightly limiting the sky coverage. Despite this drawback, the new detector offers three key advantages: 

(I) The quantum efficiency (QE) at wavelengths longer than 640\,nm is higher than for the old detector. The QE peaks at $\ge$94\% around 700\,nm and is still about 50\% at 1000\,nm, dropping sharply towards the Si band gap at 1070\,nm (see Figure~\ref{fig:QE}). During commissioning, we measured photometric zero points in imaging mode for LDSS-3C and compared them to recent measurements with the old detector system. We find only a slight decrease in Sloan $g$ (400--550\,nm) of $-0.05$\,mag, but improved zero points for Sloan $r$ (550--680\,nm) of $+0.1$\,mag, Sloan $i$ (680--830\,nm) of $+0.4$\,mag, and Sloan $z$ ($\ge$830\,nm) of $+1.5$\,mag. These numbers are fully consistent with spectroscopic measurements of standard stars, folded with the Sloan filter curves. The on-sky performance is considerably better than what was expected from the QE curves of the two CCDs. We attribute the difference, particularly the almost at par performance in the blue, to a BBAR coated dewar window in the new detector system with a reflectivity of $R<1.2$\% from 370--1000\,nm (vs.\,a MgF$_2$ coated window in the old dewar) and some degradation in performance of the old CCD after almost 10 years of continuous service.

(II) The thickness of the device greatly suppresses fringing. In fact, fringing patterns cannot be observed with the new CCD, while the old CCD showed fringing patterns rewards of 640\,nm, reaching a relative amplitude of up to 10--12\% between 840 and 980~nm. 

(III) The new CCD can be used for efficient nod-and-shuffle observations to support faint object spectroscopy. For this mode, the detector system can be manually rotated by $90\degrees$ to orient the dispersion direction with the rows of the CCD. While limiting the spectral grasp due to the limited size of the detector, this orientation allows effective charge shuffling in the spatial direction during observations. Shuffling amplitudes can be either a few arcseconds (several tens of pixels) along a short slit to accommodate very high surface densities of optical spectra \citep[see, e.g.,][]{BH01}, or 1/3 of the CCD column length in a macro nod-and-shuffle mode to improve sky subtraction for long exposures. 

Alongside the detector system, the readout electronics of LDSS-3 were exchanged with a scaled-down version of the system used for DECam \citep{Shaw12}. This system is based on a modified version of the Monsoon controller developed by CTIO. PanView \citep{Ashe02}, a LabVIEW-based astronomical instrument control system developed by Marco Bonati at CTIO, is used to control the detector and interfaces with the existing LDSS-3 instrument GUI, which was developed by Christoph Birk at Carnegie Observatories.

The full-well depth of the CCD, as determined by standardized testing procedures at Fermilab, is $\ge$200,000\,e$^-$ and it is linear within 1\% to at least 175,000\,e$^-$. To optimally use the 16\,bit depth of the analog-to-digital converter (ADC), inverse gain values are typically limited to $\le$ 2.5\,e$^-$\,ADU$^{-1}$. The dark current of the CCD is 35\,e$^-$\,hr$^{-1}$ at its operating temperature of 173\,K. Horizontal and vertical charge trap inefficiencies are below 2.5$\times10^{-5}$ and 1.1$\times10^{-5}$, respectively. The CCD has three defective columns and $\sim$6,300 bad pixels (mainly non-linear pixels), bringing the total bad pixel count to 0.19\% of the active area.  We find no amplifier crosstalk for unsaturated sources and give an upper limit of 0.02\% for potential crosstalk intensity. Saturated sources show electronic crosstalk of about 20--30\,e$^-$ when the ADC is driven into saturation. Higher levels of crosstalk (up to 100\,e$^-$) and negative crosstalk (i.e., signal levels below the bias level) are observed for badly saturated sources when both the ADC range and the full well are exceeded. 

Due to the smaller size of the CCD and two amplifiers, the fast read mode of the camera offers a 28\,sec readout, more than 2.5$\times$ faster than the old CCD. The inverse gain in this mode is 1.5 (1.8)\,e$^-$\,ADU$^{-1}$ for the left (right) amplifier and the read noise is as low as 5.5\,e$^-$ rms. A slow read mode can ideally bring the read noise down to 3\,e$^-$ rms., at the cost of a 152\,sec readout time and a low inverse gain of 0.16 (0.19)\,e$^-$\,ADU$^{-1}$, which limits the effective full-well depth to 10,000\,e$^-$. A turbo read mode is offered as well, mainly for fast target acquisition or bright sources where high duty cycles are important.  In this mode, the inverse gain is 2.7 (3.1)\,e$^-$\,ADU$^{-1}$ for the left (right) amplifier and the read noise is 8\,e$^-$ rms. 

Unfortunately, all modes still suffer from increased read noise due to electronic interference from another piece of LDSS-3 hardware. This interference increases the read noise at times up to 5\,e$^-$ rms for the slow read mode and up to 6.5\,e$^-$ rms for the fast read mode.  The shape and frequency of the noise interference pattern changes with time but is stable on the timescale of a single frame and FFT filtering techniques can be applied to suppress it. The removal of the physical source of the interference from the instrument is planned for a future appointment. We note, however, that the typical sky background between the OH lines at the resolving power of LDSS-3C, is $\ge$ 400\,e$^-$\,hr$^{-1}$ for the `VPH red' grism (quarter moon over the horizon and 60$\degr$ away from pointing) and $\ge$ 1200\,e$^-$\,hr$^{-1}$ at the end of the night for the low dispersion `VPH all' grism (quarter moon under the horizon and 70$\degr$ away from pointing). This shows that the sky background noise is still substantially larger than the read noise in the red optical and near infrared. Even for the darkest sky conditions in the bluest parts of the `VPH blue' grism (short of 450\,nm, where the sky background falls below 200\,e$^-$\,hr$^{-1}$), a read noise of 5\,e$^-$ rms is still lower than the sky background noise after 10 minutes of integration time.

A thick, fully depleted CCD has a number of `quirks', not readily apparent in thinner devices, mainly owing to the extreme geometrical aspect ratio of its pixels and its high substrate voltage (40\,V): 

(I) As discovered at Fermilab, these CCDs can actually be damaged by overexposing them with extended sources at flux levels exceeding 2,000,000\,e$^-$\,sec$^{-1}$ (10 full-well depths). At these levels, charge can get physically trapped in the CCD, permanently altering the electric fields across the device and greatly diminishing the subsequent full-well depth. Exposure to the full moon or bright sky must thus be avoided and sky flats taken with great care. 

(II) The strongly distorted electric fields at the edges of the device render 20 -- 30 pixels around the outer perimeter useless (a.k.a. 'picture frame effect'). 

(III) The single exposure time is practically limited to $\sim$30\,min for a number of reasons. (a) The sensitivity of the CCD to cosmic rays and Compton electrons from local background radiation is greatly increased compared to a thinned device. This leads to a high number of trails that might require removal strategies normally used for space-born detectors, e.g., median filtering multiple (n$\ge$3) exposures. (b) The low effective full-well depth of only 10,000\,e$^-$ in slow readout mode (normally used for long exposures) leads to saturation of the sky emission lines for the lowest dispersion mode (`VPH all' grism) after $\sim$30\,min under average conditions. (c) Charge build-up in one of the bad columns reaches a level of saturation where charge gets sucked to the edge of the frame and starts to bleed along rows. It also drives the ADC into saturation and affects the readout of neighboring columns across the frame, greatly increasing the difficulty of proper background subtraction. 

Other effects known or expected for thick CCDs, e.g., notable charge diffusion, resistivity variations in the bulk Si \citep[a.k.a.\,`tree-ring pattern', see, e.g.,][]{Astier15}, or wavelength dependent defocus from the increased absorption depth of infrared photons, are not present in LDSS-3C.

%%%%%%%%%%%%%%%%%%%%%%%%%%%%%%%%%%%%%%%%%%%%%%%%%%%%%%%%%%%%%%%%%%%%%%%%%%%%%%%
\section{OBSERVATIONS AND DATA ANALYSIS}
\label{sec:obs}
%%%%%%%%%%%%%%%%%%%%%%%%%%%%%%%%%%%%%%%%%%%%%%%%%%%%%%%%%%%%%%%%%%%%%%%%%%%%%%%

\subsection{Magellan/LDSS-3C}

\subsubsection{LDSS-3C Observation}

Using the multi-object technique developed to probe exoplanet atmospheres \citep{Bean2010}, we observed the primary transit of HAT-P-26b on the night of 16 April 2015 for nearly 5 hours (02:53 -- 07:49 UT, airmass = 1.48 $\rightarrow$ 1.19 $\rightarrow$ 1.54).  Within that time, we acquired 502 science frames using 20 second integrations.  To minimize readout times ($\sim$15 seconds), we utilized LDSS-3C's turbo read mode with low gain (2.7 and 3.1 e\sp{-}/ADU) and applied 2$\times$2 pixel binning.  Overall, we achieved a duty cycle of $\sim$57\%.

Our science masks utilized three, 12{\arcsec}-wide slits for observations of our target star (HAT-P-26, V~=~11.8) and the two comparison stars (V~=~11.1, 12.5).  Our calibration masks used corresponding slits that were only 1{\arcsec} wide to make a precise wavelength determination using He, Ne, and Ar lamps.  All of the slits were 60{\arcsec} long.  Using the VPH-Red grism, we acquired spectra between 0.7 and 1.0 {\microns}.  Unfortunately, the spectra from the brighter comparison star were too close to the detector edge to make reliable measurements; therefore, our results rely exclusively on atmospheric corrections from the fainter star.  We confirm that this device produces no measurable fringing in the red optical.

We tested inserting the OG590 filter to remove flux contribution from the blue edge of the second order spectra; however, the filter introduced clearly-visible ghosting that overlapped with the background regions of our science targets.  Because we did not detect flux from higher order spectra in the science frames, we elected to remove the filter during the acquisition of science and calibration data.

\subsubsection{LDSS-3C Reduction}

Our spectral reduction, extraction and calibration pipeline is custom software that produces multi-wavelength, systematics-corrected light curves from which we derive wavelength-dependent transit depths with uncertainties.  In \citet{Stevenson2014a}, we describe our methodology in detail.  Here, we discuss the specifics relating to our LDSS-3C observation.

The LDSS-3C detector has a large overscan region (256$\times$4096 unbinned pixel) that we use for a frame-by-frame bias correction.  Using the science mask, we acquired 20 spectroscopic flat frames prior to the observing run.  During the reduction process, we stack the set of images to form a single master flat frame then apply it uniformly to all of the science frames.  Spectroscopic flat frames using the calibration mask do not produce as good of a fit.  In additional to determining the wavelength scale, we us the calibration mask to correct for the slit tilt.

To generate an adequate model of the background flux, we apply 2$\times$2 upsampling to each frame, align pixel rows to correct for the slit tilt, mask regions containing the spectra, and fit a quadratic polynomial to the remaining pixels on a column-by-column basis.  To perform background subtraction, we reverse the alignment process, downsample to the original resolution, and subtract the background model from the unaltered data.  We then perform optimal extraction using an algorithm based on \citet{Horne1986}.

We note that the wavelength calibration results for HAT-P-26 exhibit a small offset relative to those from the two comparison stars.  We discovered this inconsistency when plotting the stellar spectra {\vs} wavelength and discerning a misalignment in the stellar spectra.  Such discrepancies can occur if all of the stars are not perfectly centered within the 12{\arcsec} slits.  To account for this translation, we manually add 0.0014 {\microns} to the wavelength calibration results for HAT-P-26.

\subsubsection{LDSS-3C White Light Curve Fits}

To correct for the observed flux variations caused by fluctuations in Earth's atmosphere, we divide the HAT-P-26 light curve by the one good comparison star.  We fit the white light curve (0.7125 -- 1.0000 {\microns}) using both transit and systematics model components.  The first utilizes a \citet{Mandel2002} transit model with free parameters for the planet-star radius ratio ($R\sb{p}/R\sb{\star}$), cosine of the inclination ($\cos i$), and semi-major axis ($a/R\sb{\star}$); and fixed quadratic limb-darkening parameters \citep[0.40792, 0.16960,][]{Claret2000} derived from stellar Kurucz models \citep{Kurucz2004}.  For the systematics component, we test various combinations of linear, quadratic, and exponential ramp functions with and without an airmass correction term.

Using the Bayesian Information Criterion \citep[BIC,][]{Liddle2008} to select the best systematics model component, our final analytic model takes the form:
\begin{equation}
\label{eqn:full}
F(t) = F\sb{\rm s}T(t)R(t)S(a),
\end{equation}
\noindent where \math{F(t)} is the measured flux at time $t$; \math{F\sb{\rm s}} is the out-of-transit system flux; \math{T(t)} is the primary-transit model component with unity out-of-transit flux; \math{R(t) = 1 + r\sb{2}(t-t\sb{0})} is the time-dependent linear model component with a fixed offset, $t\sb{0}$, and free parameter, $r\sb{2}$; and $S(a) = 1 + aA$ is the correction term at airmass, $A$.

To find the best-fit solution to any of the light curves described in this work, we use a Levenberg-Marquardt minimizer on all free parameters pertaining to that fit.  Our Differential-Evolution Markov Chain algorithm \citep[DEMC,][]{terBraak2008} estimates parameter uncertainties.  The scatter in the residuals from our LDSS-3C white light curve fit is dominated by white noise.  Table \ref{tab:transitparams} lists our best-fit physical parameters with 1$\sigma$ uncertainties.

\begin{table}[tb]
\centering
\caption{\label{tab:transitparams} 
LDSS-3C White Light Curve Transit Parameters}
\begin{tabular}{ccc}
    \hline
    \hline      
    Parameter               & Value         & Uncertainty   \\
    \hline
    Transit Midpoint (MJD\tablenotemark{a}) 
                            & 7129.72248    & 0.00017       \\
    $R\sb{P}/R\sb{\star}$   & 0.0694        & 0.0010        \\
    $\cos i$                & 0.047         & 0.008         \\
    $a/R\sb{\star}$         & 11.8          & 0.6           \\
    \hline
\end{tabular}
\tablenotetext{1}{MJD = BJD\sb{TDB} - 2,450,000}
\end{table}

\subsubsection{LDSS-3C Spectroscopic Light Curve Fits}

\begin{figure}[tb]
\centering
\includegraphics[width=1.0\linewidth]{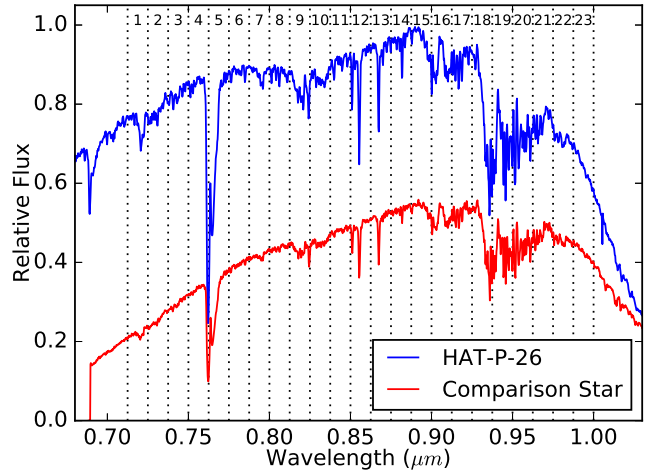}
\caption{\label{fig:H26Spec}{
LDSS-3C spectra of HAT-P-26 (blue) and the good comparison star (red).  Vertical dotted lines indicate the edges of the 23 spectrophotometric channels.
}}
\end{figure}

As part of our testing procedure, we generate spectroscopic light curves and find best-fit solutions at three different resolutions: 0.05, 0.025, and 0.0125 {\microns}.  The transmission spectra exhibit similar features at all resolutions.  For our final analysis, we select the highest resolution light curves with 23 spectroscopic bins (or channels) from 0.7125 to 1.0000 {\microns} (see Figure \ref{fig:H26Spec}).

Because we are primarily interested in the relative transit depth uncertainties, we fix the wavelength-independent parameters (transit midpoint, $\cos i$, and $a/R\sb{\star}$) to the best-fit white light curve values.  We also fix the quadratic limb-darkening parameters to those derived from our stellar Kurucz model \citep[$\log M/H = 0.0$, $T_{eff}=5000$~K, $\log g = 4.5$,][]{Kurucz2004}.

We apply the Divide White technique \citep{Stevenson2014a} to remove the wavelength-independent systematics.  To account for the wavelength-dependent systematics, each spectroscopic channel requires an airmass correction and at least a linear function in time.  Two of the channels (5 and 6) require a rising exponential plus linear model component, \math{R(t) = 1 + e\sp{-r\sb{0}t + r\sb{1}} + r\sb{2}(t-t\sb{0})} where $r\sb{0}$ -- $r\sb{2}$ are free parameters, to account for the observed ramps at the start of the night.  In each case, we use the BIC to determine the best-fitting model.  In lieu of using an exponential function for eight other channels (1, 7, 8, 13, 16, 18, 19, and 20) with comparatively weak ramps, we trim the first 23.6 (and in one case, 47.2) minutes from each of the affected datasets.  Trimming all of the datasets has no effect on our conclusions.  If this exponential ramp occurs in other LDSS-3C datasets, future observations may benefit from a 30 minute stabilization period.

\begin{figure}[tb]
\centering
\includegraphics[width=1.0\linewidth]{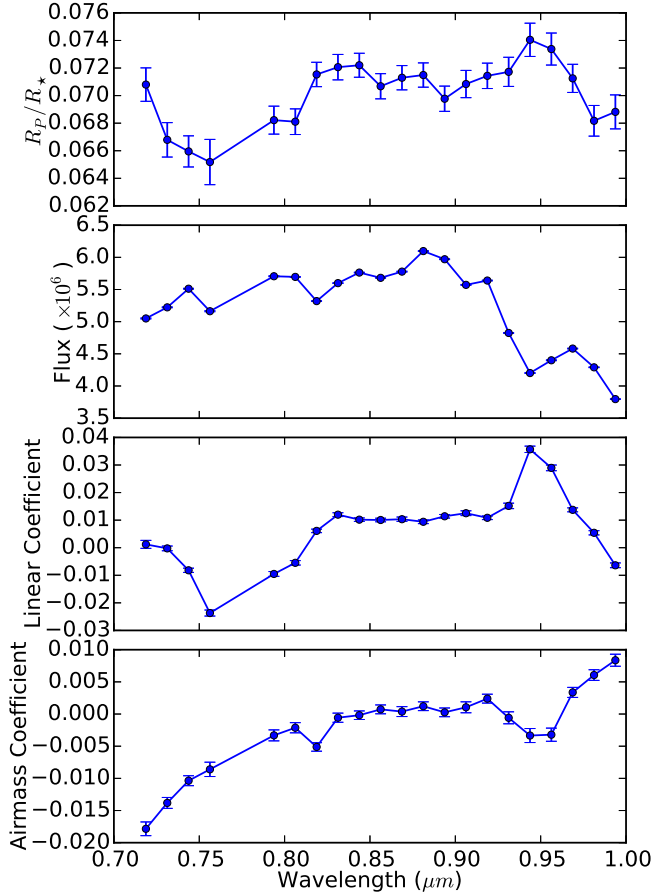}
\caption{\label{fig:FreeParams}{
Wavelength-dependent free parameters with 1$\sigma$ uncertainties for the LDSS-3C data.  Channels 5 and 6 use a more complicated rising exponential plus linear model component and are not included in this comparison.  The planet-to-star radius ratio and linear coefficient parameters are positively correlated but, upon further investigation, are not causally connected.
}}
\end{figure}

In addition to applying the above technique, we test several other methods and compare the resulting transit depths.  The first fits for the linear limb darkening coefficient, the second uses the standard technique of dividing the light curve of each spectrophotometric channel by that of its corresponding comparison star, and the third applies Gaussian Processes \citep{Ambikasaran2014} to determine the best fit.  In all three cases we find that the best-fit transit depths are within 1$\sigma$ of our final values.

Figure \ref{fig:FreeParams} displays the wavelength-dependent free parameters for the channels that utilize only a linear systematic function in time.  Values for the planet-to-star radius ratio and linear coefficient are positively correlated ($\rho=0.9$).  We investigate this further by using a shared (wavelength-independent) linear coefficient in a joint fit, but achieve visibly poor fits and uncorrelated changes in radius ratio values for the channels that exhibit strong deviations from the mean.  This confirms that these two parameters are not degenerate.  We conclude that, although the planet-to-star radius ratio and linear coefficient are correlated, they are not causally connected.  Also, because the source of the correlation is unknown, we cannot rule out the possibility that an unknown systematic may be biasing these two parameters.  Additional observations with LDSS-3C or another instrument could help to determine if this is an issue.

\begin{table}[tb]
\centering
\caption{\label{tab:depths} 
Best-Fit Transit Depths}
\begin{tabular}{cccr@{\,{\pm}\,}lc}
    \hline
    \hline      
    Channel & Wavelength          & RMS   & \mctc{Transit Depth}  & $\times$Expected   \\
            & ({\microns})        & (ppm) & \mctc{(\%)}     & Noise            \\
    \hline
    1       & 0.7125 -- 0.7250    & 1055  & 0.501   & 0.017 & 2.34       \\
    2       & 0.7250 -- 0.7375    & 1102  & 0.446   & 0.017 & 2.48       \\
    3       & 0.7375 -- 0.7500    &  968  & 0.435   & 0.015 & 2.25       \\
    4       & 0.7500 -- 0.7625    & 1406  & 0.425   & 0.021 & 3.16       \\
    5       & 0.7625 -- 0.7750    & 1363  & 0.480   & 0.025 & 2.94       \\
    6       & 0.7750 -- 0.7875    &  898  & 0.481   & 0.017 & 2.13       \\
    7       & 0.7875 -- 0.8000    &  880  & 0.465   & 0.014 & 2.08       \\
    8       & 0.8000 -- 0.8125    &  801  & 0.464   & 0.013 & 1.90       \\
    9       & 0.8125 -- 0.8250    &  806  & 0.512   & 0.013 & 1.85       \\
    10      & 0.8250 -- 0.8375    &  858  & 0.519   & 0.013 & 2.01       \\
    11      & 0.8375 -- 0.8500    &  820  & 0.521   & 0.013 & 1.94       \\
    12      & 0.8500 -- 0.8625    &  833  & 0.500   & 0.013 & 1.97       \\
    13      & 0.8625 -- 0.8750    &  780  & 0.508   & 0.013 & 1.86       \\
    14      & 0.8750 -- 0.8875    &  836  & 0.511   & 0.013 & 2.03       \\
    15      & 0.8875 -- 0.9000    &  829  & 0.487   & 0.013 & 1.99       \\
    16      & 0.9000 -- 0.9125    &  859  & 0.502   & 0.014 & 2.02       \\
    17      & 0.9125 -- 0.9250    &  831  & 0.510   & 0.013 & 1.97       \\
    18      & 0.9250 -- 0.9375    &  962  & 0.514   & 0.015 & 2.08       \\
    19      & 0.9375 -- 0.9500    & 1138  & 0.548   & 0.018 & 2.27       \\
    20      & 0.9500 -- 0.9625    & 1045  & 0.538   & 0.017 & 2.15       \\
    21      & 0.9625 -- 0.9750    &  923  & 0.508   & 0.015 & 1.95       \\
    22      & 0.9750 -- 0.9875    &  973  & 0.465   & 0.015 & 1.98       \\
    23      & 0.9875 -- 1.0000    & 1120  & 0.474   & 0.017 & 2.11       \\
    \hline
    IRAC1   & 3.6                 & 4734  & 0.522   & 0.012 & 1.03       \\
    IRAC2   & 4.5                 & 6125  & 0.559   & 0.029 & 1.20       \\
    \hline
\end{tabular}
\end{table}

\begin{figure*}[tb]
\centering
\includegraphics[width=0.49\linewidth]{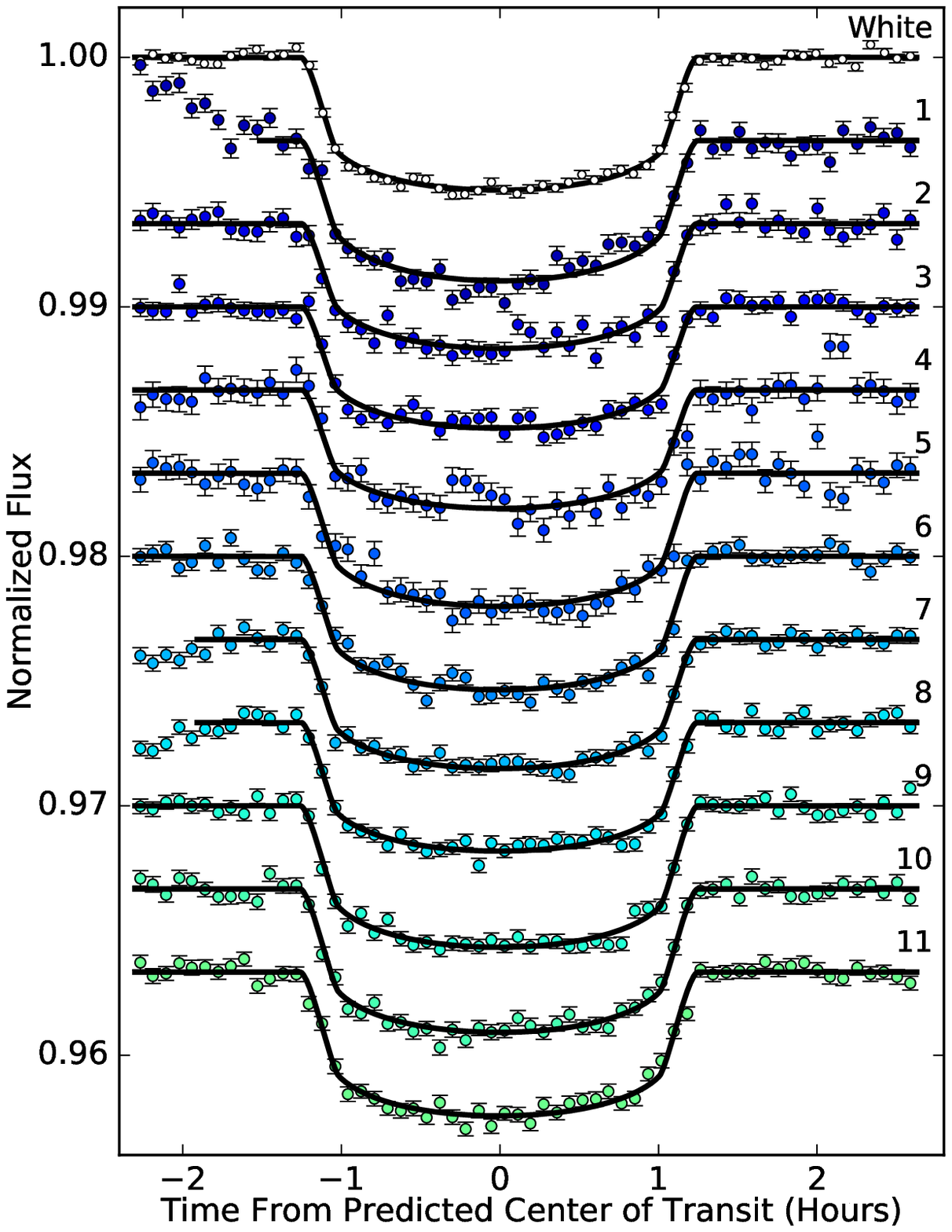}
\includegraphics[width=0.49\linewidth]{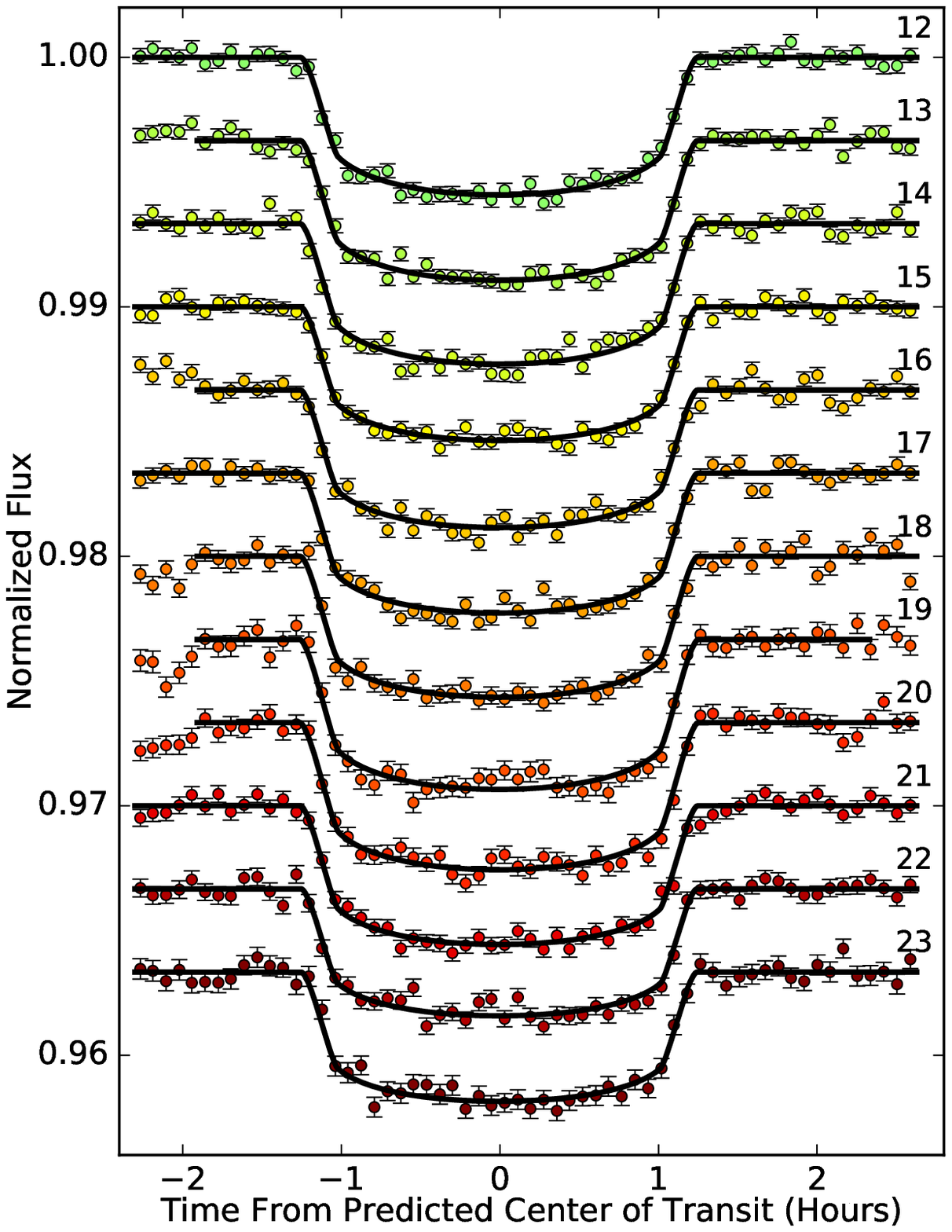}
\caption{\label{fig:ldss3ctransits}{
LDSS-3C primary transit of HAT-P-26b in 23 spectrophotometric channels.  The colored symbols represent binned data that are normalized with respect to the system flux.  Also included is the band-integrate (white) light curve (white symbols).  The solid black lines indicate the best fit models.  Light curves are vertically offset for clarity.  The channel number for each spectroscopic light curve is listed on the right.
}}
\end{figure*}

Figure \ref{fig:ldss3ctransits} displays both the white and spectroscopic light curves after removing the modeled systematics.  No channel exhibits significant time-correlated noise, as determined by plotting the theoretical and actual rms values over a range of time scales.  We provide light curve rms values, best-fit transit depths, and precisions normalized to those expected at the photon limit for HAT-P-26b in Table \ref{tab:depths}.  Channels 4 and 5 overlap Earth's O\sb{2} feature (as seen in Figure \ref{fig:H26Spec}) and its variability likely contributes to the increased scatter seen in these channels.

%%%%%%%%%%%%%%%%%%%%%%%%%%%%%%%%%%%%%%%%%%%%%%%%%%%%%%%%%%%
%  SPITZER                                                %
%%%%%%%%%%%%%%%%%%%%%%%%%%%%%%%%%%%%%%%%%%%%%%%%%%%%%%%%%%%
\subsection{Spitzer Space Telescope}

\subsubsection{Spitzer Observations and Reduction}

As part of Program 90092 (PI: Jean-Michel D\'esert), {\em Spitzer} observed two transits of HAT-P-26b during the warm mission.  The InfraRed Array Camera \citep[IRAC,][]{IRAC} acquired broadband photometric data in subarray mode at 3.6~{\microns} (09 September 2013) and 4.5~{\microns} (23 April 2013).  Both observations commenced with 30 minutes of settling time to mitigate spacecraft drift.  The science observations lasted $\sim$4.5 hours and acquired 8,128 frames in each channel utilizing two second frame times. 

To reduce the data, we used the Photometry for Orbits, Eclipses, and Transits (POET) pipeline \citep{Campo2011,Stevenson2011,Cubillos2013}.  For this analysis, POET flagged bad pixels using a double-iteration, 4$\sigma$ filter at each pixel column in stacks of 64 subarray frames, determined image centers from a 2D Gaussian fit \citep{Lust2014}, and applied $5\times$ interpolated aperture photometry \citep{Harrington2007} over a range of aperture sizes in 0.25 pixel increments.

\subsection{Spitzer Light-Curve Systematics and Fits}

At 3.6 and 4.5 {\microns}, the dominant systematic is the intra-pixel sensitivity effect, wherein small, $\sim$0.01 pixel pointing variations can cause measurable changes in the observed flux.  Numerous groups have recently devised new solutions that effectively account for this systematic without relying on the measured centroid positions \citep[e.g.,][]{Deming2015, Evans2015, Morello2015}; however, in this work, we apply the robust and proven method of Bilinearly-Interpolated Subpixel Sensitivity (BLISS) mapping \citep{Stevenson2011} to model the position-dependent systematics.  The {\em Spitzer} light curves also exhibit a weak, time-dependent trend that we model using a linear function.

\begin{table}[tb]
\centering
\caption{\label{tab:Spitzerparams} 
Best {\em Spitzer} Analysis and Transit Parameters}
\begin{tabular}{ccc}
    \hline
    \hline      
    Parameter               & 3.6 {\microns}\tablenotemark{a}    
                                                & 4.5 {\microns}\tablenotemark{a}    \\
    \hline
    Aperture Size (pixels)  & 2.50              & 2.50              \\
    Ramp Model              & Linear            & Linear            \\
    Intra-pixel Model       & BLISS             & BLISS             \\
    System Flux ($\mu$Jy)   & 44060(10)         & 27088(4)          \\
    Transit Midpoint (MJD\tablenotemark{b}) 
                            & 6545.3622(3)      & 6405.6237(9)      \\
    $R\sb{P}/R\sb{\star}$   & 0.0725(8)         & 0.0748(19)        \\
    $\chi^{2}_{\nu}$        & 1.04              & 1.20              \\
    \hline
\end{tabular}
\tablenotetext{1}{Parentheses indicate 1$\sigma$ uncertainties in the least significant digit(s).}
\tablenotetext{2}{MJD = BJD\sb{TDB} - 2,450,000.}
\end{table}

\begin{figure}[tb]
\centering
\includegraphics[width=1.0\linewidth]{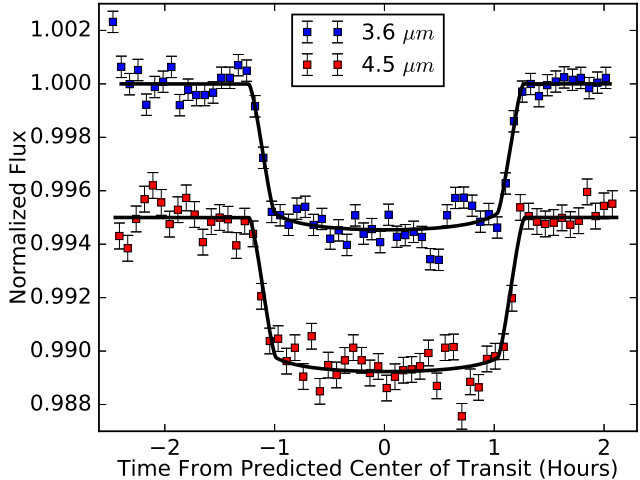}
\caption{\label{fig:spitzertransits}{
{\em Spitzer} primary transits of HAT-P-26b at 3.6 and 4.5 {\microns}.  The colored symbols represent binned data that are normalized with respect to the system flux.  The solid black curves indicate the best fit models.  Light curves are offset vertically for clarity.
}}
\end{figure}

In fitting the shape of the primary transit, we fix $\cos i$ and $a/R\sb{\star}$ to the values reported in Table \ref{tab:transitparams}.  We adopt a quadratic stellar limb-darkening model with fixed values of (0.10260, 0.15726) at 3.6 {\microns} and (0.09509, 0.11323) at 4.5 {\microns} \citep{Kurucz2004}.  Both datasets exhibit signs of time-correlated noise, which we account for by applying the wavelet analysis described by \citet{Carter2009b}.  Table \ref{tab:Spitzerparams} lists our final reduction parameters and best-fit transit values with uncertainties in parentheses.  Figure \ref{fig:spitzertransits} depicts the normalized, systematics-removed, 3.6 and 4.5 {\micron} {\em Spitzer} light curves of HAT-P-26b.  Although time-correlated noise is evident in both IRAC channels, it only increases the 4.5 {\micron} transit depth uncertainty (by a factor of two).

%%%%%%%%%%%%%%%%%%%%%%%%%%%%%%%%%%%%%%%%%%%%%%%%%%%%%%%%%%%%%%%%%%%%%%%%%%%%%%%
\section{RESULTS AND DISCUSSION}
\label{sec:results}
%%%%%%%%%%%%%%%%%%%%%%%%%%%%%%%%%%%%%%%%%%%%%%%%%%%%%%%%%%%%%%%%%%%%%%%%%%%%%%%

\subsection{Orbital Analysis}

\begin{figure}[tb]
\centering
\includegraphics[width=1.\linewidth]{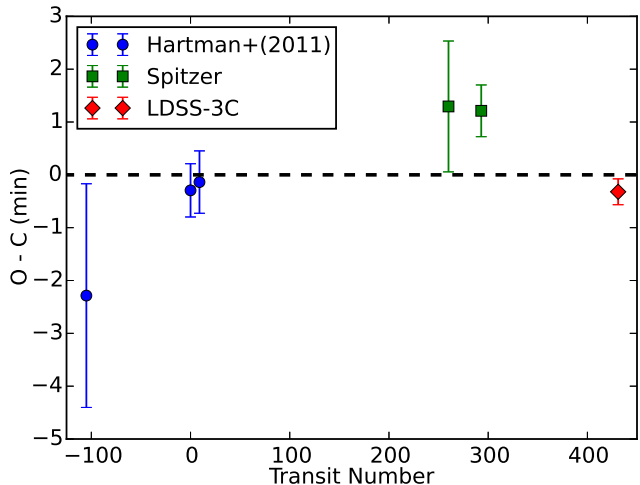}
\caption{\label{fig:OC}{
Observed minus calculated transit times of HAT-P-26b.  With our LDSS-3C and {\em Spitzer} data (red diamonds and green squares, respectively), we compute an updated orbital period and transit ephemeris.
}}
\end{figure}

Using the transit times published by \citet{Hartman2011} and the values derived from our LDSS-3C and {\em Spitzer} analyses, we recompute HAT-P-26b's transit ephemeris ($t$\sb{0}~= 2455304.65218(25)~BJD\sb{TDB}) and orbital period ($p$~= 4.2345023(7)~days).  Using these new values, we plot the planet's observed minus calculated (O-C) transit times in Figure~\ref{fig:OC}.  We note a weak curvature in the residuals that requires precise follow-up observations to confirm.

\subsection{Atmospheric Models}

For comparison to HAT-P-26b's measured transmission spectrum, we generate a series of representative atmospheric forward models using the methods described in detail by \citet{Line2013a, Line2015, Swain2014, Kreidberg2014, Kreidberg2015}.  Given the quality of our fits (see Table \ref{tab:chi2}), we acknowledge that a full atmospheric retrieval is unwarranted and would likely produce fallacious constraints.  The transmission forward models consider different metallicities, the presence of obscuring clouds, and the strength of the potassium feature centered at 0.768 {\microns}. 

\begin{figure}[tb]
\centering
\includegraphics[width=1.0\linewidth]{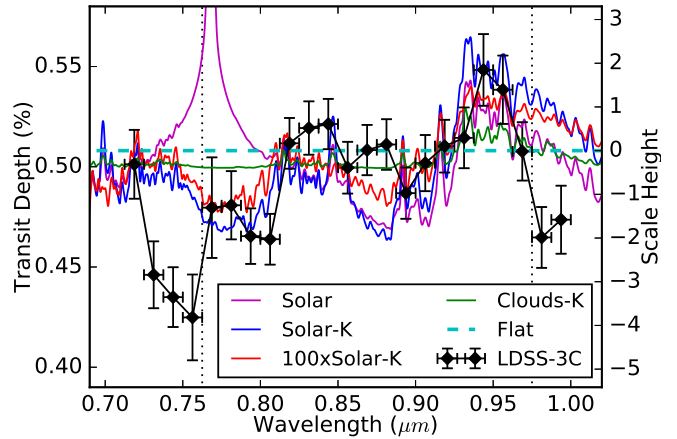}
\caption{\label{fig:ldss3Spec}{
LDSS-3C transmission spectrum of HAT-P-26b.  Representative atmospheric models without potassium (blue, red, and green solid curves) achieve better fits to the measured LDSS-3C spectrum (black diamonds) than those with potassium (magenta solid curve).  A flat model (dashed cyan line) also achieves a reasonable fit between 0.7625 and 0.9750~{\microns} (vertical dotted lines).  The source of the five systematically low transit depths (relative to our models) is unknown. 
}}
\end{figure}

\begin{figure*}[tb]
\centering
\includegraphics[width=1.0\linewidth]{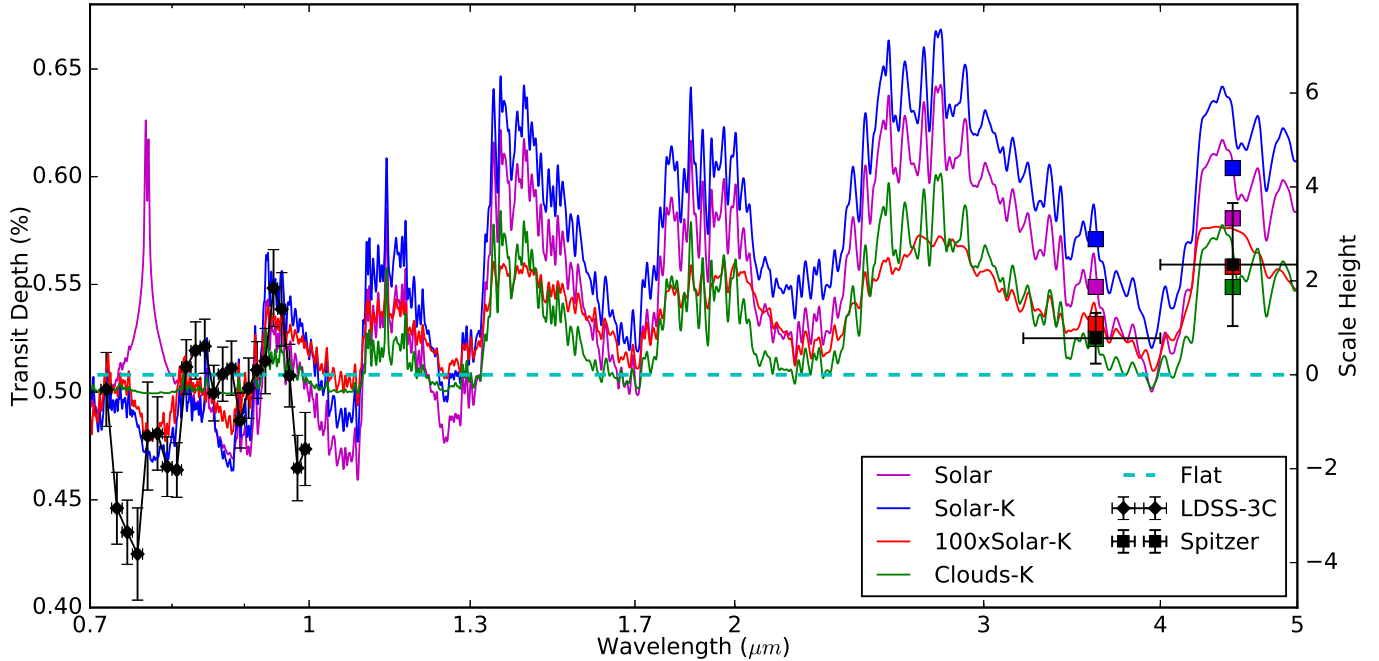}
\caption{\label{fig:fullspec}{
LDSS-3C and {\em Spitzer} transmission spectrum of HAT-P-26b.  The description is the same as Figure~\ref{fig:ldss3Spec}, but includes two {\em Spitzer} data points (black squares).  The combined spectrum favors either a high-metallicity atmosphere (red solid curve) or a solar-metallicity atmosphere with a 10~mbar cloud deck (green solid curve).  A flat spectrum (cyan dashed line) cannot be ruled out at high confidence.
}}
\end{figure*}

A description of our atmospheric methods is as follows.  Our model solves the equations described in \citet{Brown2001} and \citet{Tinetti2012}.  To generate the spectra presented in Figures \ref{fig:ldss3Spec} and \ref{fig:fullspec}, we specify the thermal structure, composition, cloud properties, and 10 bar radius.  Transit transmission spectra are largely insensitive to the shape of the thermal profile; therefore, for simplicity, we assume a thermal structure consistent with radiative equilibrium at full redistribution with zero albedo using the analytic formulae given by \citet{Line2013a}.  \citet{Guillot2010} describe the thermal structure parameters (resulting in a skin temperature of 900~K) and we determine molecular abundances by assuming thermochemical equilibrium along the temperature-pressure profile at the given metallicity using the NASA Chemical Equilibrium with Applications Code \citep{Gordon1994}.  We fix the elemental ratios (with the exception of potassium) to solar and include as opacity sources: H\sb{2}-H\sb{2}/He collision-induced absorption, H\sb{2}O, CH\sb{4}, CO, CO\sb{2}, NH\sb{3}, Na, K, TiO, VO, C\sb{2}H\sb{2}, HCN, H\sb{2}S, and FeH, with the relevant line lists provided by \citet{Freedman2014}.  Molecular Rayleigh scattering is parameterized using the prescription and values described by \citet{LecavelierDesEtangs2008}.  The cloud is modeled as an opaque gray absorber, of which the atmospheric transmittance is zero below the cloud top pressure.  We adjust the 10 bar radius to match the vertical offset in the spectra.  The models assume a surface gravity of $\log g = 2.66$.

In Figure \ref{fig:ldss3Spec}, we compare our models to the measured LDSS-3C spectrum.  Our representative atmospheric models without potassium achieve good fits to the data between 0.7625 and 0.9750~{\microns} (Channels 5 -- 21); however, outside of this wavelength range there are five outliers that are systematically low and do not agree with any of our models.  Channel 4 overlaps with Earth's strongest O\sb{2} absorption band, which contributes excessive noise to the spectroscopic light curve and has been shown to meaningfully alter the measured transit depth \citep{Parviainen2015}.  The other affected channels do not coincide with Earth's telluric lines; therefore, it's unclear if the source of these inconsistent transit depths is physical or instrumental in nature and, if physical, extrasolar or local in origin.  The measured transit depths are independent of our choice of limb-darkening models and the number of spectrophotometric channels.  Without additional information to identify the source, we recommend performing follow-up observations of HAT-P-26b using one or more instruments at overlapping wavelengths.

The addition of our two {\em Spitzer} data points allows us to rule out a cloud-free, solar-metallicity atmosphere (see Figure \ref{fig:fullspec}).  However, this additional constraint still cannot differentiate between a high-metallicity, cloud-free atmosphere or a solar-metallicity atmosphere with a 10~mbar cloud deck because both scenarios similarly reduce the size of the spectral features.  A precise {\em HST}/WFC3 or blue optical transmission spectrum should be able to distinguish between these competing scenarios.  If HAT-P-26b follows the mass-metallicity relation seen in our own solar system \citep{Kreidberg2014b} then, given a mass of 0.0586 M\sb{J}, one would expect an atmospheric metallicity of 60 -- 100$\times$ solar.  Therefore, a cloud-free, high-metallicity atmosphere is consistent with current expectations.

\begin{table}[tb]
\centering
\caption{\label{tab:chi2} 
Comparison of Atmospheric Models}
\begin{tabular}{ccccc}
    \hline
    \hline      
    Label       & Metallicity   & Potassium & Cloud Level   & $\chi^2_{\nu}$\tablenotemark{a}\\
    \hline
    100xSolar-K & 100$\times$Solar & No     & None          & 1.5           \\
    Clouds-K    & Solar         & No        & 10 mbar       & 1.7           \\
    Clouds      & Solar         & Yes       & 10 mbar       & 2.2           \\
    100xSolar   & 100$\times$Solar & Yes    & None          & 2.4           \\
    Flat        & --            & No        & --            & 2.6           \\
    Solar-K     & Solar         & No        & None          & 3.5           \\
    Solar       & Solar         & Yes       & None          & 5.0           \\
    \hline
\end{tabular}
\tablenotetext{1}{Includes two {\em Spitzer} points and 17 LDSS-3C points from 0.7625 to 0.9750~{\microns}.}
\end{table}

Table \ref{tab:chi2} lists the $\chi^2_{\nu}$ values of each representative model when including channels 5 -- 21 from LDSS-3C and both {\em Spitzer} points.  Using all of the LDSS-3C channels in our calculation increases the $\chi^2_{\nu}$ values by 2.  Although all of the forward models with $\chi^2_{\nu} < 3$ remain statistically-plausible scenarios, the evidence currently favors the detection of water and a lack of potassium, which condenses out at $\sim$1000~K at 1~bar.  Since $T$\sb{eq} = 990~K for HAT-P-26b, the condensation of potassium into KCl is a plausible scenario.  As a test, we reduce the temperature in our forward models until potassium completely condenses out, but find that this does not occur until $T < 800$~K.  HAT-P-26b's terminator could reach this temperature if the planet has a relatively large albedo and/or poor day-night heat redistribution.

\subsection{Comparison to Other Planets}

To place our results into a broader context, we compare HAT-P-26b to other exoplanets with similar temperatures and surface gravities.  These two factors are thought to most strongly influence the production of obscuring clouds and hazes; therefore, selecting planets with similar values provides for an unbiased comparison.  We identify two planets comparable to HAT-P-26b: HAT-P-12b ($T\sb{eq} = 960$~K, $\log g = 2.75$~dex) and HAT-P-19b ($T\sb{eq} = 1010$~K, $\log g = 2.75$~dex).

\citet{Line2013c} observed a single transit of HAT-P-12b using {\em HST}/WFC3.  They report no evidence for water absorption in the transmission spectrum from 1.1--1.7 {\microns} and conclude that the data are best described by a high-altitude cloud model.  However, because the data were taken in the staring mode (before the advent of the spatial scan), the transit depth uncertainties are relatively large and a spectrum containing a truncated water feature cannot be ruled out.

Using the OSIRIS spectrograph at the Gran Telescopio Canarias, \citet{Mallonn2015a} observed a single transit of HAT-P-19b.  They report transit depths in the range of 562 -- 767~nm that are consistent with a flat spectrum.  However, this wavelength region does not contain water absorption features and their precision is insufficient to rule out the presence of a pressure-broadened sodium feature.

Although evidence for water in the LDSS-3C transmission spectrum of HAT-P-26b cannot be corroborated by similar detections from exoplanets with comparable equilibrium temperatures and surface gravities, there is also no strong precedence against such a possibility.  Fortunately, pending {\em HST}/STIS and WFC3 observations of HAT-P-26b (GO 14110, PI: David Sing; GO 14260, PI: Drake Deming) will provide the necessary precision to make a definitive detection and potentially distinguish between a high-metallicity atmosphere or a solar-metallicity atmosphere with a 10~mbar cloud deck.

%%%%%%%%%%%%%%%%%%%%%%%%%%%%%%%%%%%%%%%%%%%%%%%%%%%%%%%%%%%%%%%%%%%%%%%%%%%%%%%
\section{CONCLUSIONS}
\label{sec:concl}
%%%%%%%%%%%%%%%%%%%%%%%%%%%%%%%%%%%%%%%%%%%%%%%%%%%%%%%%%%%%%%%%%%%%%%%%%%%%%%%

Here we present the first results from the recently-upgraded LDSS-3C instrument on Magellan.  The detector's enhanced sensitivity and suppressed fringing in the red optical enables it to effectively search for the spectroscopic signature of water in an exoplanet atmosphere from the ground.

Targeting the Neptune-mass planet HAT-P-26b, we find tentative evidence for water and a lack of potassium in its transmission spectrum.  Since the data are not precise enough to warrant a full atmospheric retrieval, we compare the measured spectrum to several representative forward models.  We conclude that HAT-P-26b is likely to have a high-metallicity, cloud-free atmosphere or a solar metallicity atmosphere with cloud deck at $\sim$10~mbar.  Although more high-precision data are needed to break this degeneracy, a 100$\times$Solar-metallicity atmosphere is consistent with expectations based on Uranus and Neptune's similar metallicities.  Alternatively, since HAT-P-26b has a similar equilibrium temperature and surface gravity to that of HAT-P-12b (which exhibits a featureless transmission spectrum that is likely due to high-altitude clouds), it is conceivable that the smaller HAT-P-26b simply has lower-altitude clouds and, thus, a metallicity that is closer to solar.  The presence of multi-scale-height spectral features in our data suggests that future observations at higher precision could reveal the planet's atmospheric composition and metallicity.

\acknowledgments

We thank all contributors to the LDSS-3C project, particularly our colleagues at Fermilab (Steve Chappa, Greg Derylo, Tom Diehl, Juan Estrada, Brenna Flaugher, Lee Scott, Terri Shaw, Walter Stuermer, Donna Kubik, and Kevin Kuk), at Carnegie and Las Campanas Observatories (Christoph Birk, Alan Uomoto, and David Osip), and at The University of Chicago (Mike Gladders and Josh Frieman). We are deeply indebted to Marco Bonati at CTIO for tailoring PanView to LDSS-3C and thank him for his countless hours of work on this project.  We thank contributors to SciPy, Matplotlib, and the Python Programming Language, the free and open-source community, the NASA Astrophysics Data System, and the JPL Solar System Dynamics group for software and services.  Funding for this work has been provided by NASA grant NNX13AJ16G.  K.B.S. recognizes support from the Sagan Fellowship Program, supported by NASA and administered by the NASA Exoplanet Science Institute (NExScI).  J.L.B. acknowledges support from the Alfred P.~Sloan and David and Lucile Packard Foundations.

\bibliography{ms}

\end{document}

%% file: kbsmacs.tex
\newcommand\degree{\degr}
\newcommand\degrees\degree
\newcommand\vs{\em vs.}

\DeclareSymbolFont{UPM}{U}{eur}{m}{n}
\DeclareMathSymbol{\umu}{0}{UPM}{"16}
\let\oldumu=\umu
\renewcommand\umu{\ifmmode\oldumu\else\math{\oldumu}\fi}
\newcommand\micro{\umu}
\renewcommand\micron{\micro m}
\newcommand\microns \micron

\renewcommand\arcsec[0]{$^{\prime\prime}$}

\let\oldsim=\sim
\renewcommand\sim{\ifmmode\oldsim\else\math{\oldsim}\fi}
\let\oldpm=\pm
\renewcommand\pm{\ifmmode\oldpm\else\math{\oldpm}\fi}
\newcommand\by{\ifmmode\times\else\math{\times}\fi}

\newbox{\wdbox}
\renewcommand\c{\setbox\wdbox=\hbox{,}\hspace{\wd\wdbox}}
\renewcommand\i{\setbox\wdbox=\hbox{i}\hspace{\wd\wdbox}}

\newcount\timect
\newcount\hourct
\newcount\minct
\newcommand\now{\timect=\time \divide\timect by 60
         \hourct=\timect \multiply\hourct by 60
         \minct=\time \advance\minct by -\hourct
         \number\timect:\ifnum \minct < 10 0\fi\number\minct}

\newcommand\mctc{\multicolumn{2}{c}}

\catcode`@=11

\newcommand\comment[1]{}
\newcommand\commenton{\catcode`\%=14}
\newcommand\commentoff{\catcode`\%=12}

\renewcommand\math[1]{$#1$}
\newcommand\mathshifton{\catcode`\$=3}
\newcommand\mathshiftoff{\catcode`\$=12}

\comment{the backslash is necessary}

\comment{alignment tab}

\let\atab=&
\newcommand\atabon{\catcode`\&=4}
\newcommand\ataboff{\catcode`\&=12}

\let\oldmsp=\sp
\let\oldmsb=\sb
\def\sp#1{\ifmmode
           \oldmsp{#1}%
         \else\strut\raise.85ex\hbox{\scriptsize #1}\fi}
\def\sb#1{\ifmmode
           \oldmsb{#1}%
         \else\strut\raise-.54ex\hbox{\scriptsize #1}\fi}
\newbox\@sp
\newbox\@sb
\def\sbp#1#2{\ifmmode%
           \oldmsb{#1}\oldmsp{#2}%
         \else
           \setbox\@sb=\hbox{\sb{#1}}%
           \setbox\@sp=\hbox{\sp{#2}}%
           \rlap{\copy\@sb}\copy\@sp
           \ifdim \wd\@sb >\wd\@sp
             \hskip -\wd\@sp \hskip \wd\@sb
           \fi
        \fi}
\def\msp#1{\ifmmode
           \oldmsp{#1}
         \else \math{\oldmsp{#1}}\fi}
\def\msb#1{\ifmmode
           \oldmsb{#1}
         \else \math{\oldmsb{#1}}\fi}
\def\supon{\catcode`\^=7}
\def\supoff{\catcode`\^=12}
\def\subon{\catcode`\_=8}
\def\suboff{\catcode`\_=12}
\def\supsubon{\supon \subon}
\def\supsuboff{\supoff \suboff}

\newcommand\actcharon{\catcode`\~=13}
\newcommand\actcharoff{\catcode`\~=12}

\newcommand\paramon{\catcode`\#=6}
\newcommand\paramoff{\catcode`\#=12}

\comment{And now to turn us totally on and off...}

\newcommand\reservedcharson{\commenton \mathshifton \atabon \supsubon \actcharon
	\paramon}

\newcommand\reservedcharsoff{\commentoff \mathshiftoff \ataboff
	\supsuboff \actcharoff \paramoff}

\catcode`@=12
\reservedcharsoff

\reservedcharson

\comment{ Must have ONLY ONE of these... trust these macros, they work

}

\newcommand{\squishlist}{
 \begin{list}{$\bullet$}
  { \setlength{\itemsep}{1pt}
     \setlength{\parsep}{0pt}
     \setlength{\topsep}{3pt}
     \setlength{\partopsep}{0pt}
     \setlength{\leftmargin}{2.0em}
     \setlength{\labelwidth}{1.5em}
     \setlength{\labelsep}{0.5em} } }

\newcommand{\squishend}{
  \end{list}  }

\reservedcharson
\actcharon